\documentclass[preprint,aps]{revtex4}

\usepackage{latexsym}
\usepackage[dvips]{graphicx}
\usepackage{epsf}
\def\singlespace {\smallskipamount=3.75pt plus1pt minus1pt
                  \medskipamount=7.5pt plus2pt minus2pt
                  \bigskipamount=15pt plus4pt minus4pt
                  \normalbaselineskip=12pt plus0pt minus0pt
                  \normallineskip=1pt
                  \normallineskiplimit=0pt
                  \jot=3.75pt
                  {\def\smallskip {\vskip\smallskipamount}}
                  {\def\medskip   {\vskip\medskipamount}}
                  {\def\bigskip   {\vskip\bigskipamount}}
                  {\setbox\strutbox=\hbox{\vrule
                    height10.5pt depth4.5pt width 0pt}}
                  \parskip 7.5pt
                  \normalbaselines}
\def\middlespace {\smallskipamount=5.625pt plus1.5pt minus1.5pt
                  \medskipamount=11.25pt plus3pt minus3pt
                  \bigskipamount=22.5pt plus6pt minus6pt
                  \normalbaselineskip=22.5pt plus0pt minus0pt
                  \normallineskip=1pt
                  \normallineskiplimit=0pt
                  \jot=5.625pt
                  {\def\smallskip {\vskip\smallskipamount}}
                  {\def\medskip   {\vskip\medskipamount}}
                  {\def\bigskip   {\vskip\bigskipamount}}
                  {\setbox\strutbox=\hbox{\vrule
                    height15.75pt depth6.75pt width 0pt}}
                  \parskip 11.25pt
                  \normalbaselines}
\def\doublespace {\smallskipamount=7.5pt plus2pt minus2pt
                  \medskipamount=15pt plus4pt minus4pt
                  \bigskipamount=30pt plus8pt minus8pt
                  \normalbaselineskip=30pt plus0pt minus0pt
                  \normallineskip=2pt
                  \normallineskiplimit=0pt
                  \jot=7.5pt
                  {\def\smallskip {\vskip\smallskipamount}}
                  {\def\medskip   {\vskip\medskipamount}}
                  {\def\bigskip   {\vskip\bigskipamount}}
                  {\setbox\strutbox=\hbox{\vrule
                    height21.0pt depth9.0pt width 0pt}}
                  \parskip 15.0pt
                  \normalbaselines}

\def\frac#1#2{\textstyle{{{#1} \over {#2}}}}

\def\lsim{\mathrel{\rlap{\lower4pt\hbox{\hskip1pt$\sim$}}
    \raise1pt\hbox{$<$}}}
\def\gsim{\mathrel{\rlap{\lower4pt\hbox{\hskip1pt$\sim$}}
    \raise1pt\hbox{$>$}}}

\def\etal {{\it et al.}}

\newcommand{\beq}{\begin{equation}}
\newcommand{\eeq}{\end{equation}}
\newcommand{\bea}{\begin{eqnarray}}
\newcommand{\eea}{\end{eqnarray}}

\begin{document}
\preprint{
\hfill$\vcenter{\hbox{\bf IUHET-512}}$  }
\title{\vspace*{.75in}
Possible Equilibria of Interacting Dark Energy Models}

\author{Micheal S. Berger\footnote{Electronic address:berger@indiana.edu}
 and Hamed Shojaei\footnote{Electronic address:seshojae@indiana.edu}}

\affiliation{Physics Department, Indiana University, Bloomington, IN 47405, USA}

\date{\today}

\begin{abstract}
Interacting dark energy and the holographic principle offer a possible
way of addressing the cosmic coincidence problem as well as accounting for the
size of the dark energy component. The equilibrium points
of the Friedmann equations which govern the evolution behavior of dark energy,
matter, and curvature components can determine the qualitative behavior of
the cosmological models. These possible equilibria and their behavior are
examined in a general framework, and some illustrative examples are presented.
\end{abstract}

\maketitle

\section{Introduction}
The study of type Ia supernovae indicates that we live in an
accelerating Universe. This acceleration can be described by adding
another component to the cosmic inventory that
behaves in a similar way to the famous cosmological constant. Observations are consistent with about 70\% of the density of the
nearly flat universe is in the form of dark energy with the remaining
30\% of ingredients is composed of matter, baryonic or non-baryonic.
During the expansion of the Universe, one component tends to replace
another in a small number of e-foldings, so the current closeness of these two
densities is remarkable. This is known as the cosmic coincidence
problem~\cite{Steinhardt:1997}. A perhaps associated question concerns why the dark energy
density is so small compared to the other physical scales in particle
physics~\cite{Weinberg:1988cp}. These issues have led to the consideration of models that are
more general than adding a simple cosmological constant.  One
hope is that a model based on a simple, single physical principle can
account for both the coincidence and the observed size of
the dark energy density.

A class of models that has been employed to explain the
coincidence problem involves a nonzero interaction between dark energy
and matter. In these theories, dark energy converts into matter in such a
way as to create an equilibrium balance at late times. The choice of an
appropriate interaction can effect a smooth and gradual transition
from a matter-dominated (decelerating) Universe to a dark energy
dominated (accelerating) one. These models 
lack a fundamental explanation for the source of the interaction. They are nevertheless an attractive way to investigate, at least
at the phenomenological level, ways to account for the observed cosmological
quantities such as the dark energy density parameter $\Omega_\Lambda
= 0.7$ and its equation of state $w_\Lambda \approx -1$. If dark
energy decays into matter, it is possible to imagine an equilibrium
situation in which the tendency of the dark energy component to
dominate as the Universe expands, is balanced by its decay into
matter. These models represent deviations from, and are more general
than, the model with a cosmological constant
($\Lambda$CDM).

The problem of the scale (size) of the dark energy component has led
to the consideration of the holographic principle. This idea offers a
possible mechanism for tying the scale of the dark energy to the
geometric average of the Planck scale and the size of the Universe.
The authors previously have used interacting dark energy in the scheme
of holographic principle to find a relation between the equations
governing the dark energy and matter densities in a flat
universe~\cite{Berger:2006db}. A holographic principle relates the
density of dark energy to some horizon. Some examples are the apparent Hubble,
future event horizon or particle horizon~\cite{Fischler:1998st,Bousso:1999xy,Huang:2004wt,Myung:2005sv,Myung:2005pw,Zhang:2005hs,Kim:2005at,Horava:2000tb,Setare:2007eq,Gong:2004cb}.
It has been previously noted that, without any interaction between dark
energy and matter, only the future event horizon gives results
in agreement with observations~\cite{Li:2004rb}. However, if we introduce an
interaction, all of the above choices for the horizon might be consistent. There
exists, for example, a model in which a constant interaction with Hubble
horizon in a flat universe gives rise to the desired
properties~\cite{Berger:2006fk}.

The consideration of nonzero curvature is the next obvious
generalization~\cite{Kim:2006kk}. Although we know from the cosmic microwave
background that we live in an almost flat universe, this may not have
been the case at all times. Furthermore, the decay of dark energy to
matter involves the conversion of a component which tends to make the
universe more flat with one that has the opposite behavior. Nonzero
curvature also makes the evolution equations more involved, and one can
accommodate additional fixed point solutions.

We describe the equations that govern the
dynamics of dark energy and curvature in Section II.  Assuming that
the radiation component is negligible, these are the only two
independent density parameters out of the three (matter component,
dark energy component, and the ``curvature component'') related by
$\Omega_m+\Omega_\Lambda=1+\Omega_k$. To solve the evolution equations
one needs to know the effective equations of state for dark energy and
matter. This requires a specification of the holographic
condition on dark energy, i.e. whether the length scale associated with the
dark energy density is the Hubble horizon, the particle horizon, the
future horizon, or some other physical scale. In addition, one needs
to specify some interaction. With these two specifications, solving
the resulting differential equations analytically is in general
impossible, and solutions have to be done numerically.
Therefore, it is desirable to arrive at some general characterization
of the types of asymptotic solutions that can be obtained. One needs
to find the steady-state solutions of the equations, where the variable parameters
are fixed for large times in the future or past. These equilibrium
solutions, and the conditions giving rise to them, are discussed in Section
III. Some of these conditions have been exploited previously in
the literature. To further clarify the main points of our results, we
describe different examples in detail in the Section IV.
It is worth mentioning that interacting models are not limited to models which employ an interaction between dark energy and matter. There exist models in which there are two components for dark energy: a variable cosmological constant, $\Lambda$, and a dynamical "Cosmon", $X$, with possible interaction between them. These models, called $\Lambda$XCDM models, are also capable of addressing the cosmological constant problem. See e.g.~\cite{Grande:2006nn,Grande:2006gb,Grande:2007wj}.

\section{Dynamical Equations for $\Omega_\Lambda$ and $\Omega_k$}

We start by assuming that the universe consists of matter and dark energy,
nonzero curvature, and with negligible radiation. The evolution equations for
dark energy and matter are,
\bea
\dot{\rho}_\Lambda + 3H(1+w_\Lambda)\rho_\Lambda&=& -Q\;, \nonumber \\
\dot{\rho}_m + 3H(1+w_m)\rho_m&=&Q\; .
\label{twocomp}
\eea
We shall refer to the quantities $w_m$ and $w_\Lambda$ as the native
equations of state.
The appearance of the same $Q$ in the two equations insures the
overall conservation of the energy-momentum tensor. Positive $Q$ can
be interpreted as a transfer from the dark energy component to the
matter component. Presumably this interaction could arise from some
microscopic mechanism and might be specified more fully when the nature of the
dark energy is known. It can
be noted, however, that the consideration of quantum fields in curved
backgrounds gives rise to particle production. However, in a cosmological
scenario based on the Friedmann-Lemaitre-Robertson-Walker (FLRW)
metric, the production rate is only at the level of the Hubble
constant (and not the geometric average of the Hubble scale and the
Planck scale). It is interesting to note that the required
physical scale of the interaction involves, like the dark energy itself, the
size of the universe. This suggests a holographic interpretation for any such
interaction.

The quantity $\rho_m$ is the matter component for which one usually
takes $w_m=0$.  The quantity $\rho_\Lambda$, the density of dark
energy can be related to some length scale via a holographic
principle. We define this length scale $L_\Lambda$ according to the
equation
\bea
\rho_\Lambda={{3c^{2}M_{Pl}^{2}}\over{8\pi
    L_\Lambda^{2}}}\;.
\eea Here $c$ is a constant of order
one (inserted here to preserve agreement with the literature) and $M_{Pl}$ is the Planck mass.
Various choices considered for this length scale are the Hubble
horizon $R_H=1/H$, the particle horizon (PH) defined by
\bea
R_{PH}=a\int\limits_{0}^{t}{{dt}\over{a}}
=a\int\limits_{0}^{a}{{da}\over{Ha^2}}\;,
\label{PH}
\eea
and the future event horizon (FH) defined by
\bea
R_{FH}=a\int\limits_{t}^{\infty}{{dt}\over{a}}
=a\int\limits_{a}^{\infty}{{da}\over{Ha^2}}\;,
\label{FH}
\eea
where $a(t)$ is the scale factor in the FLRW
metric.

The interaction can be combined with the native equations of state,
$w_\Lambda$ and $w_m$, to define effective equations of state as
\bea
\dot{\rho}_\Lambda + 3H(1+w_\Lambda^{\rm eff})\rho_\Lambda&=&0\;,
\nonumber \\
\dot{\rho}_m + 3H(1+w_m^{\rm eff})\rho_m &=& 0\;.
\label{definew}
\eea
Defining the ratio $r=\rho_m/\rho_\Lambda$ and the rate
$\Gamma =Q/\rho_\Lambda$ and assuming a pressureless matter component, for
which $w_m=0$, these effective equations of states
are
\bea
w_\Lambda ^{\rm eff}=w_\Lambda+{{\Gamma}\over {3H}}\;, \qquad
w_m ^{\rm eff}=-{1\over r}{{\Gamma}\over {3H}}\;.
\label{eos}
\eea
The density parameters
\bea
\Omega_\Lambda={{8\pi\rho_\Lambda}\over {3M_{Pl}^2H^2}}\;, \qquad
\Omega_m={{8\pi\rho_m}\over {3M_{Pl}^2H^2}}\;, \qquad \Omega_k ={k \over {H^2a^2}}\;.
\eea
are defined so that
the Friedmann equation,
\bea
H^2= {{8\pi G}\over {3}}(\rho_\Lambda+\rho_m)- {{k}\over {a^2}}\;.
\label{fried}
\eea
gives
\bea
\Omega_\Lambda +\Omega_m=1+\Omega_k\;.
\label{omegas}
\eea

The ratio $r$ is related to density parameters by
\bea
r={{1-\Omega_\Lambda+\Omega_k}\over {\Omega_\Lambda}}\;,
\eea
and its time evolution is
\bea
\dot{r}=3Hr\left [w_\Lambda-w_m+{{1+r}\over r}{{\Gamma}\over {3H}}\right ]
=3Hr\left [w_\Lambda^{\rm eff}-w_m^{\rm eff}\right ]\;.
\label{rdot}
\eea
It is evident from Eq.~(\ref{rdot}) that the condition $w_m^{\rm eff}=w_\Lambda^{\rm eff}$ involving the effective
equations of state gives
$\dot{r}=0$. When this equilibrium occurs, the ratio of dark energy and
dark matter
densities is a constant. The Friedmann equation, Eq.~(\ref{fried}),
and the time evolution of densities can be combined to find the time
evolution of Hubble parameter,
\bea
{{1}\over {H}}{{dH}\over{dx}}=-{{3}\over{2}}-{{1}\over{2}}\Omega_k
-{{3}\over{2}}w_\Lambda\Omega_\Lambda\;,
\label{Hdot}
\eea
where $x=ln(a/a_0)$ with some fixed scale factor $a_0$.
The density parameters satisfy the differential equations,
\bea
{{d\Omega_\Lambda}\over {dx}}&=& 3\Omega_\Lambda\left [{{1}\over {3}}\Omega_k
+w_\Lambda(\Omega_\Lambda - 1)-{{\Gamma}\over {3H}}\right ]\;,
\nonumber \\
{{d\Omega_k}\over {dx}}&=&\Omega_k(1+\Omega_k+3w_\Lambda\Omega_\Lambda)\;.
\label{prelim}
\eea
It was shown in Ref.~\cite{Berger:2006db} that it is sufficient to have
two physical conditions to determine the evolution of the parameters
of the universe. The possibilities include a
holographic condition on the dark energy $\rho_\Lambda$, an assumption
about the nature of dark energy and its native equation of state,
$w_\Lambda$ or an assumption for the form of interaction $Q$, or
equivalently $\Gamma$. These three specifications are related by
\bea
\Gamma=3H(-1-w_\Lambda)+2{{\dot L_\Lambda}\over{L_\Lambda}}.
\eea
This equation demonstrates that the interaction should generically be
of the same size of the Hubble parameter, and also suggests that holographic
definitions for the interaction may be useful.

Since the physical interpretation of the effective equation of
state is clear, one
can eliminate $w_\Lambda$ and $\Gamma$ in favor of
$w_\Lambda^{\rm eff}$ and $w_m^{\rm eff}$ to obtain the equations,
\bea
{{d\Omega_\Lambda}\over {dx}}&=&-3\Omega_\Lambda(1-\Omega_\Lambda)
(w_\Lambda^{\rm eff}-w_m^{\rm eff})+\Omega_k\Omega_\Lambda(1+3w_m^{\rm eff})\;,
\label{lambdadot}
\nonumber \\
{{d\Omega_k}\over{dx}}&=&3\Omega_k\Omega_\Lambda(w_\Lambda^{\rm eff}
-w_m^{\rm eff})+\Omega_k(1+\Omega_k)(1+3w_m^{\rm eff})\;.
\label{kdot}
\eea
These equations are consistent with the analysis
of Ref.~\cite{Kim:2006kk}, and with the substitution $\Omega_k=0$, one recovers the equation for the flat case
from Ref.~\cite{Berger:2006db}. The location of the fixed points
and equilibria of these coupled differential equations will determine their
asymptotic behavior. The appearance of the factors
$w_\Lambda^{\rm eff}-w_m^{\rm eff}$ and $1+3w_m^{\rm eff}$ are easy to
understand on physical grounds. The first factor merely compares whether
dark energy or matter comes to dominate as the universe expands. The
second factor compares matter to curvature
(``$w_k^{\rm eff}=-{{1}\over{3}}$''),
so it measures whether the
density of matter increases or decreases as the universe expands.

\section{Possible Equilibrium Solutions of the Evolution Equations}

Equilibrium solutions occur when the right hand sides of the
differential equations, Eqs.~(\ref{kdot}),
vanish. We wish to distinguish different kinds of equilibria. First, there are
fixed points in $\Omega_\Lambda,\Omega_k$ space. These are universal in the
sense that they will be present
unless the effective equations of state have singularities as
a density parameter vanishes. Second, some equilibria consist of one
constraint on the effective equations of state and a fixed value of
$\Omega_\Lambda$ or $\Omega_k$. This represents a fixed point which depends in
detail on the specific form of the interaction and holographic definition of the
dark energy. Finally, there is an equilibrium
solution which is
governed by just constraints on effective equation of states.

A procedure to identify these equilibria is the following: Set the
derivatives in Eqs.~(\ref{kdot}) to zero. Taking linear
combinations of the right-hand-sides, one finds solutions
\bea
\Omega_\Lambda(w_\Lambda^{\rm eff}-w_m^{\rm eff})&=&0\;, \nonumber \\
\Omega_k(1+3w_m^{\rm eff})&=&0\;.
\eea
One obtains solutions by identifying points where the density parameters vanish
and/or where the effective equations of state satisfy a certain condition, namely, $w_\Lambda^{\rm eff}=w_m^{\rm eff}$ or $1+3w_m^{\rm eff}=0$.

The behavior near these
fixed points is important in
understanding the qualitative behavior of the evolution. The fixed points can be categorized as either
repellers, attractors or  saddle points.
Consider the two differential equations in the following
form~\cite{Setare:2007we,Campos:2001pa,Campos:2001cn},
\bea
{{d\bf\Omega}\over{dx}}=\bf f(\bf\Omega)\;.
\label{bold}
\eea In this equation $\bf\Omega$ $=(\Omega_\Lambda,\Omega_k)$ and
$\bf f(\bf\Omega)$ represents the right hand sides of the two
differential equations of the evolutions of the density parameters.
The density parameter for matter, $\Omega_m$ is then given by
Eq.~(\ref{omegas}).  The behavior at a specific equilibrium point
depends on the eigenvalues of the 2-by-2 matrix
$\bf A = {{\partial\bf f}\over{\partial\bf \Omega}}$.
If the eigenvalues of this matrix
are all positive, then we have a repeller fixed point. In other words,
as the Universe expands and $x$ increases, the Universe will tend to
move away from the fixed point.  If the eigenvalues are all negative,
we have an attractor fixed point. Finally if one eigenvalue is positive and one is negative, the fixed point behaves like a saddle point.

There are three fixed points for the system of equations which occur
at certain values of the density parameters. The character of these
fixed points is easy to identify from the properties of the effective 
equations of state.  The first fixed point occurs at ${\bf
\Omega}=(0,-1)$. This is a negatively curved solution, and would be a future asymptotic state of the
Universe if both $w_m^{\rm eff}>-{{1}\over{3}}$ and $w_\Lambda^{\rm
  eff}>-{{1}\over{3}}$.  Since the Universe appears flat, we know that
it is unlikely to be a viable solution. The second fixed point occurs
at ${\bf \Omega}=(0,0)$. This corresponds to a universe filled
with only matter. This could be the final state of the Universe, provided
$w_m^{\rm eff}<-{{1}\over{3}}$ (so that the matter density grows as the
Universe expands) and that $w_\Lambda^{\rm eff}>w_m^{\rm eff}$ (so that
matter density comes to dominate over the dark energy density).
Finally, the de Sitter Universe is described by ${\bf \Omega}=(1,0)$,
a universe just filled with dark energy. This solutions pertains to
the future asymptotic state of the Universe provided $w_\Lambda^{\rm
  eff}>-{{1}\over{3}}$ (so that the matter density grows as the
universe expands) and $w_\Lambda^{\rm eff}<w_m^{\rm eff}$. The
eigenvalues of the matrix $\bf A$ are shown in Table~\ref{firsttable}.
If one replaces the effective equations of state with their native
values ($w_m=0$ for matter and $w_\Lambda=-1$ for a cosmological
constant), one recovers the usual evolution behavior associated with
the $\Lambda$CDM.  One recognizes from Table~\ref{firsttable} that the
behavior in a more general context is governed entirely by the values
of the effective equations of state at the fixed point.

It is also worth mentioning that, for specific interactions, some of
these fixed points can be eliminated. For example, if the interaction
produces a $w_m^{\rm eff}$ proportional to something sufficiently
singular like $1/\Omega_\Lambda$, then the point ${\bf \Omega}=(0,0)$
will cease to be a fixed point as can be seen in Eq.~(\ref{eos}).

\begin{table}
\begin{center}
\begin{tabular}{|c|c|} \hline
fixed point& eigenvalues of $\bf A$\\ \hline
${\bf \Omega}=(0,-1)$  &  $-(1+3w_\Lambda^{\rm eff})$\\
    $\Omega_m=0$ &$-(1+3w_m^{\rm eff})$\\ \hline
${\bf \Omega}=(0,0)$   &  $-3(w_\Lambda^{\rm eff}-w_m^{\rm eff})$\\
  $\Omega_m=1$ &$1+3w_m^{\rm eff}$\\ \hline
${\bf \Omega}=(1,0)$  &  $3(w_\Lambda^{\rm eff}-w_m^{\rm eff})$\\
  $\Omega_m=0$ &$1+3w_\Lambda^{\rm eff}$\\ \hline
\end{tabular}
\caption{Equilibria of the universal fixed points in $\bf\Omega$ space.
The signs of the eigenvalues are given by comparing the effective
equations of state to the that of the curvature component (which
is $-1/3$), or by comparing them to each other.}
\label{firsttable}
\end{center}
\end{table}

The behavior of the evolution near these fixed points
depends on the effective equations of state in
Table~\ref{firsttable}, the different possibilities for their
dynamical behaviors
are shown in Table~\ref{4table}, \ref{5table} and \ref{6table}.

\begin{table}
\begin{center}
\begin{tabular}{|c|c|c|}\hline
&$w_\Lambda^{\rm eff}< -{{1}\over{3}}$&$w_\Lambda^{\rm eff}> -{{1}\over{3}}$
\\ \hline
$w_m^{\rm eff}<-{{1}\over{3}}$&repeller&saddle point\\ \hline
$w_m^{\rm eff}>-{{1}\over{3}}$&saddle point&attractor\\ \hline
\end{tabular}
\caption{Characteristic
behavior of the fixed point ${\bf \Omega}=(0,-1)$, when $\Omega_m=0$.}
\label{4table}
\end{center}
\end{table}

\begin{table}
\begin{center}
\begin{tabular}{|c|c|c|}\hline
&$w_\Lambda^{\rm eff}< -{{1}\over{3}}$&$w_\Lambda^{\rm eff}> -{{1}\over{3}}$
\\ \hline
$w_m^{\rm eff}<-{{1}\over{3}}$&attractor if $w_\Lambda^{\rm eff}>w_m^{\rm eff}$&attractor\\
&saddle point if $w_\Lambda^{\rm eff}<w_m^{\rm eff}$&\\ \hline
$w_m^{\rm eff}>-{{1}\over{3}}$&repeller&repeller if $w_\Lambda^{\rm eff}<w_m^{\rm eff}$\\
&&saddle point if $w_\Lambda^{\rm eff}>w_m^{\rm eff}$\\ \hline
\end{tabular}
\caption{Characteristic
behavior of fixed point ${\bf \Omega}=(0,0)$, when $\Omega_m=1$.}
\label{5table}
\end{center}
\end{table}

\begin{table}
\begin{center}
\begin{tabular}{|c|c|c|}\hline
&$w_\Lambda^{\rm eff}< -{{1}\over{3}}$&$w_\Lambda^{\rm eff}> -{{1}\over{3}}$
\\ \hline
$w_m^{\rm eff}<-{{1}\over{3}}$&saddle point if $w_\Lambda^{\rm eff}>w_m^{\rm eff}$&repeller\\
&attractor if $w_\Lambda^{\rm eff}>w_m^{\rm eff}$&\\ \hline
$w_m^{\rm eff}>-{{1}\over{3}}$&attractor&saddle point if $w_\Lambda^{\rm eff}<w_m^{\rm eff}$\\
&&repeller if $w_\Lambda^{\rm eff}>w_m^{\rm eff}$\\ \hline
\end{tabular}
\caption{Characteristic b
ehavior of fixed point ${\bf \Omega}=(1,0)$, when $\Omega_m=0$.}
\label{6table}
\end{center}
\end{table}

The fixed points just discussed are the familiar ones that result from
behavior of each density component as the universe expands. There are
also fixed point, or equilibrium solutions, that can result because
the effective equations of state satisfy some constraint.  There are
two equilibrium solutions where just one of the density parameters is
determined and the other condition is a constraint on the equations of
state.  The first occurs when $\Omega_\Lambda = 0$ and $w_m^{\rm
  eff}=-{{1}\over{3}}$. This equilibrium would be possible only if the
chosen interaction gives rise to
${{\Gamma}\over{3H}}\sim{{1}\over{\Omega_\Lambda}}$ around
$\Omega_\Lambda=0$. Although this doesn't directly make the
interaction $Q$ infinite, it forces $\Gamma$ to be infinite around
$\Omega_\Lambda=0$, which might be problematic (see Eq.~(\ref{eos})).
The second equilibrium occurs when $\Omega_k = 0$ and $w_m^{\rm
  eff}=w_\Lambda^{\rm eff}$. This solution has been
employed~\cite{Berger:2006db},~\cite{Kim:2005at},~\cite{Zimdahl:2000zm}
to represent the final asymptotic state of the universe, and it has to
be an attractor to be a suitable candidate.
This implies that $\dot{r}=0$ from Eq.~(\ref{rdot}) and the density
parameters remain constant, provided this is a stable equilibrium as
determined by the eigenvalues at the fixed point.  The current epoch
of the universe may be nearing such a fixed point.  On the other hand,
it is also possible that the current state of the universe is a
transient one in the sense that behavior near the equilibrium point
behaves like a saddle point. The eigenvalues of $\bf A$ are listed in
Table~\ref{secondtable}.

\begin{table}
\begin{center}
\begin{tabular}{|c|c|c|} \hline
fixed density parameter & constraint on effective EOS& eigenvalues
of $\bf A$\\ \hline
$\Omega_\Lambda =0$&$w_m^{\rm eff}=-{{1}\over{3}}$&$-(1+3w_\Lambda^{\rm eff})$\\
&&$3\Omega_k(1+\Omega_k){{\partial w_m^{\rm eff}}
\over {\partial \Omega_k}}$\\ \hline
$\Omega_k = 0$& $w_m^{\rm eff}=w_\Lambda^{\rm eff}$&
$-3\Omega_\Lambda(1-\Omega_\Lambda)({{\partial w_\Lambda^{\rm eff}}
\over {\partial \Omega_\Lambda}}-{{\partial w_m^{\rm eff}}
\over {\partial \Omega_\Lambda}})$\\
&&$1+3w_m^{\rm eff}$\\ \hline
\end{tabular}
\caption{Equilibria specified by one fixed density parameter and one
constraint.}
\label{secondtable}
\end{center}
\end{table}

The last equilibrium condition occurs when the constraint
$w_m^{\rm eff}=w_\Lambda^{\rm eff}=-{{1}\over{3}}$ holds. The equilibrium
solution is easy to understand as both effective equations of state
scale like curvature, so that the dark energy density and matter
density parameters are constant.
It is harder to see the behavior of this equilibrium as the
derivative matrix, $\bf A$, includes complicated function of the
derivatives of the effective equations of state.
An example of a solution satisfying these constraints was presented in
Ref.~\cite{Berger:2006fk}. There the model consisted of a holographic length scale
set equal to the Hubble parameter, and a constant interaction $\Gamma$.

These examples indicate the large variety of behavior that can result
with fixed points.  The behavior of the equations at these other
equilibria is more involved than the universal fixed points in
Table~\ref{firsttable}, as the eigenvalues of $\bf{A}$ depend on the
derivatives of the effective equations of state as well. Some examples
will be considered in next section to illustrate the different
possibilities.

\section{Examples}

In this section the behavior of the density parameters will be illustrated
with some choices for the interaction.
As mentioned earlier, two conditions are required to completely specify the equations. If one of the conditions is determined by a
holographic principle, it will restrict the possible behaviors of the
equilibria. If one chooses the length scale entering the
equation of density of dark energy to be the future event horizon,
defined by Eq.~(\ref{FH}), then
it can be shown that the effective equation of state for dark energy can be
written as ~\cite{Setare:2006wh}
\bea
w_\Lambda^{\rm eff}=-{{1}\over{3}}-{{2}
\over{3}}\sqrt{{{\Omega_\Lambda}\over{c^2}}-\Omega_k}\;.
\label{FHeos}
\eea
On the other hand, if the length scale is chosen to be the
particle horizon, defined by Eq.~(\ref{PH}), with a similar
procedure one can show that
\bea
w_\Lambda^{\rm
  eff}=-{{1}\over{3}}+{{2}\over{3}}
\sqrt{{{\Omega_\Lambda}\over{c^2}}-\Omega_k}\;.
\label{PHeos}
\eea
In general in the FH case,
$w_\Lambda^{\rm eff}\leq -{{1}\over{3}}$, and for the PH case,
$w_\Lambda^{\rm eff}\geq -{{1}\over{3}}$. Therefore, after
specification of this condition, the behavior of the fixed points will
be determined by the behavior of $w_m^{\rm eff}$, which is in turn
determined by the form of the interaction.

\begin{figure}
\centerline{
 \includegraphics[width=3.50in]{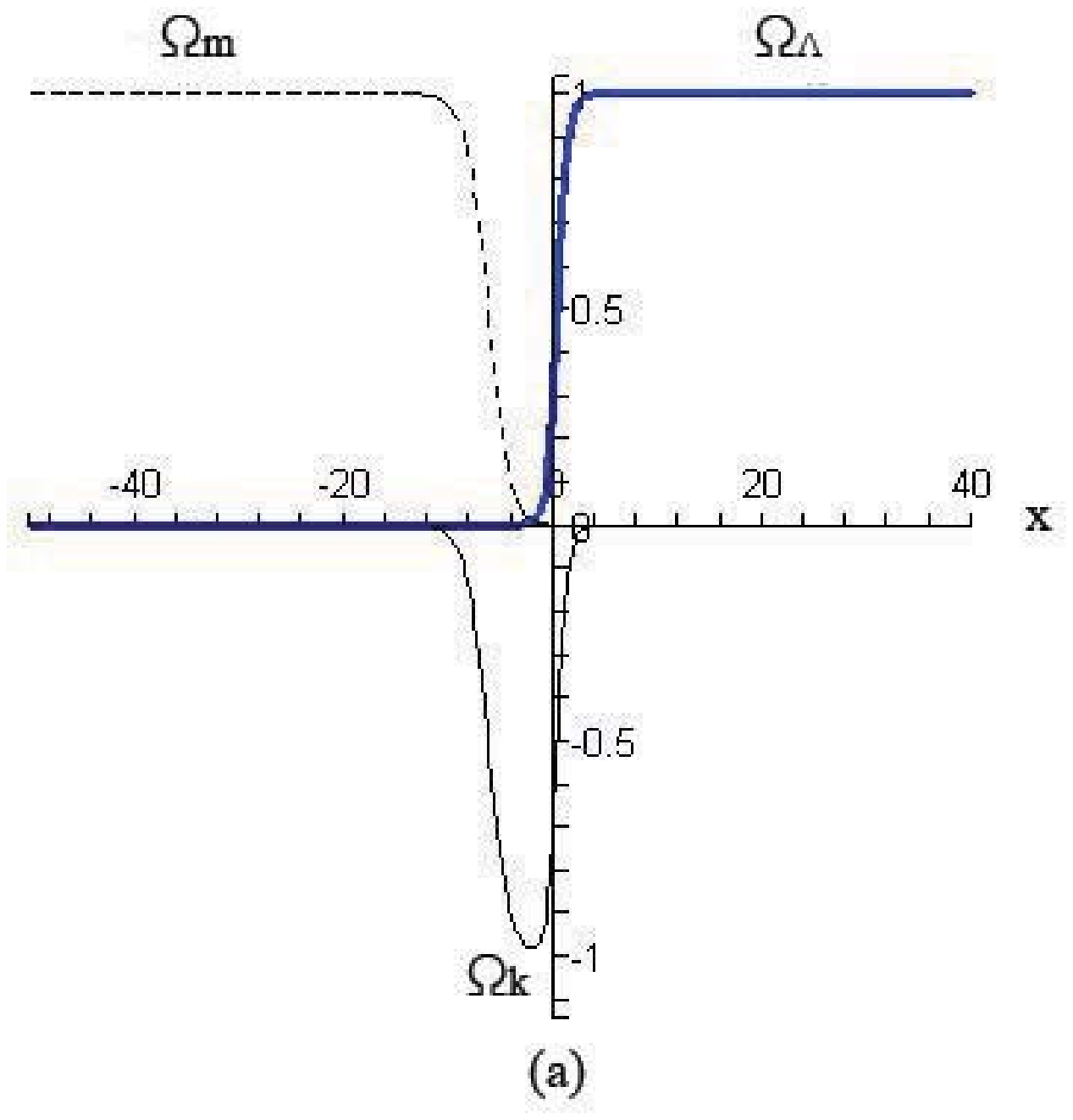}
 \includegraphics[width=3.50in]{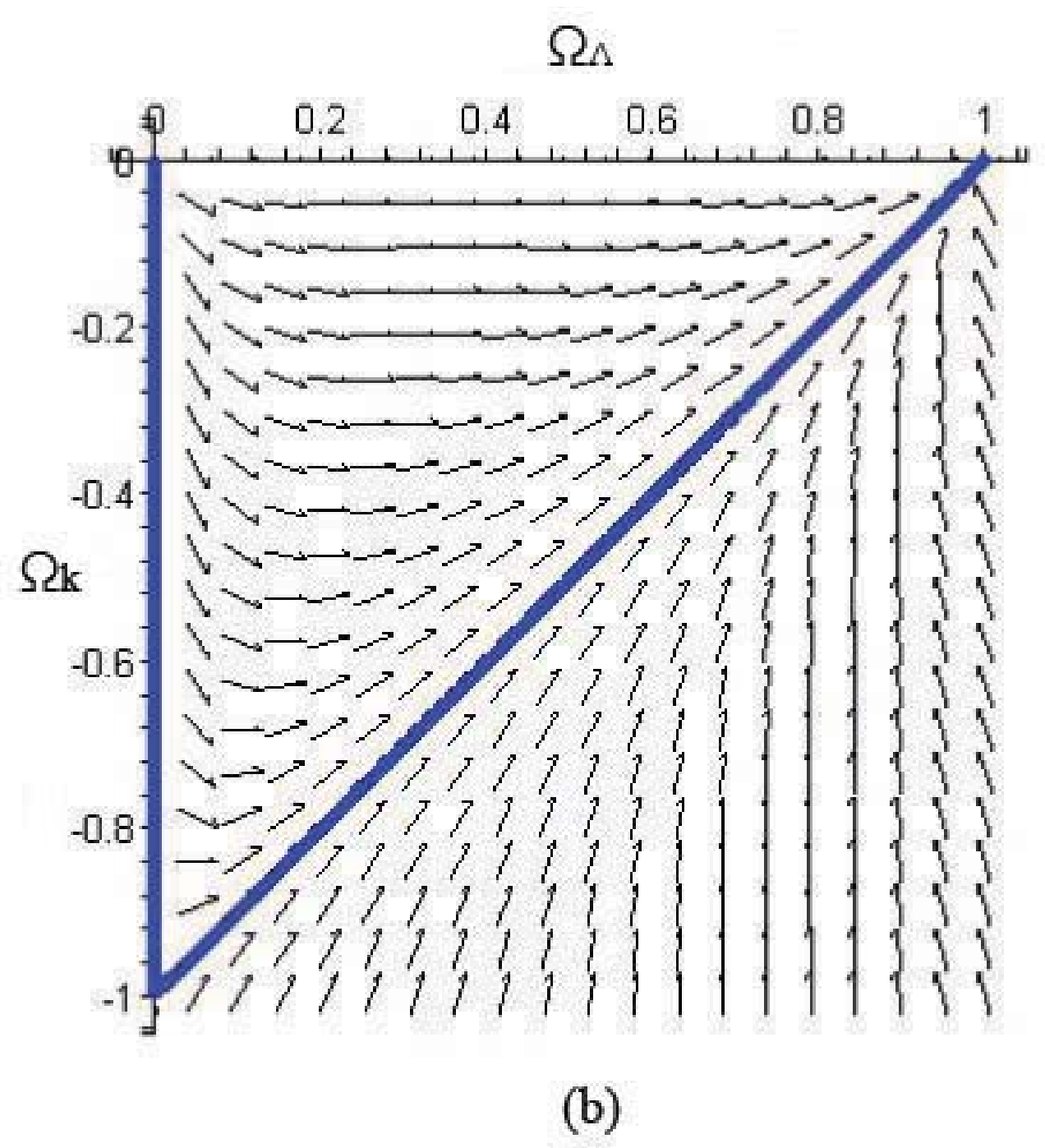}
}
\caption{(a) Density parameters in a model with no interaction.
(b) The associated flow diagram in the $\Omega_\Lambda, \Omega_k$ plane.
The solid line represents
the solution for the initial condition assumed in (a).}
\label{graph}
\end{figure}

For all the examples in this section, the holographic length scale,
entering in the equation of dark energy density, will be taken to be
the future event horizon, $L_\Lambda=R_{FH}$ defined in
Eq.~(\ref{FH}).  The effective equation of state for dark energy is
given by Eq.~(\ref{FHeos}). What remains is a
specification of the form of the interaction. The simplest choice is no
interaction at all, for which $w_m^{\rm eff}=0$. In this case, there
exist just the three equilibrium points listed in
Table~\ref{firsttable}.  The behavior of the fixed points can be
obtained by looking at Tables~\ref{4table},
\ref{5table} and \ref{6table}, considering the fact that $w_m^{\rm
  eff}=0$ and $w_\Lambda^{\rm eff}<-{{1}\over{3}}$. The fixed points
at ${\bf \Omega}=(0,0)$, ${\bf \Omega}=(0,-1)$, and ${\bf
  \Omega}=(1,0)$ are a repeller, a saddle point, and an attractor
respectively\footnote{For the particle horizon (PH), the behavior at
  ${\bf \Omega}=(0,0)$ and ${\bf \Omega}=(1,0)$ depends on whether
  $w_\Lambda^{\rm eff}<w_m^{\rm eff}$ or not.}.  The behavior of the
density parameters are shown in Fig.~\ref{graph}, where an initial state is chosen so that the Universe is
filled with matter, while dark energy and curvature are very
small. The interaction drives the Universe toward a final equilibrium
in which the Universe is filled with dark energy. The
Universe experiences a transition toward and then away from a saddle point
where it is negatively curved. The physical interpretation of this behavior
is similar to but not the same as $\Lambda$CDM: matter tends to create curvature
in the universe, but eventually the dark energy component comes to dominate
and reduces the curvature.

\bigskip
\bigskip
The next example employs an interaction between dark energy and dark matter and illustrates the appearance of a new equilibrium point. In terms of $\Gamma$, the interaction is
\bea
{{\Gamma}\over{3H}}={{b^2(1+\Omega_k)}\over{\Omega_\Lambda}^n}\;,
\label{usefulinteraction}
\eea
which has been used in some recent works to explain the cosmic coincidence problem\cite{Kim:2005at,Wang:2005jx}. When $n=1$, the interaction term $Q=\rho_\Lambda\Gamma$ is just dependent on $H$, the Hubble parameter. The density parameters, acting under this interaction, approach an equilibrium solution similar to their observed values at the present time. The effective equation of state for matter is
\bea
w_m^{\rm eff}={{-b^2(1+\Omega_k)\Omega_\Lambda^{1-n}}
\over{1+\Omega_k-\Omega_\Lambda}}.
\eea
 It can be seen from Eqs.~(\ref{kdot}) and the definition of $w_m^{\rm eff}$ that this interaction eliminates $\bf\Omega$ = (1,0) as an equilibrium. However, with an appropriate choice for $b^2$, it produces a new attractor equilibrium at $\bf\Omega$ = (0.7,0). This new fixed
point arises from an equilibrium specified by
$w_m^{\rm eff}=w_\Lambda^{\rm eff}$ as in Table~\ref{secondtable}.
This attractor
can be employed to represent the future asymptotic state of the Universe. Although there is no theoretical justification for this interaction, it turns out to be very useful in understanding how introducing interactions can change the behavior of the density parameters. The initial values for density parameters don't have any effect on their asymptotic behavior. Initial values, though, dictate how the density parameters evolve before reaching their final asymptotic value and how near they approach the saddle point-like equilibria.
\bigskip
\begin{figure}
\centerline{
 \mbox{\includegraphics[width=2.50in]{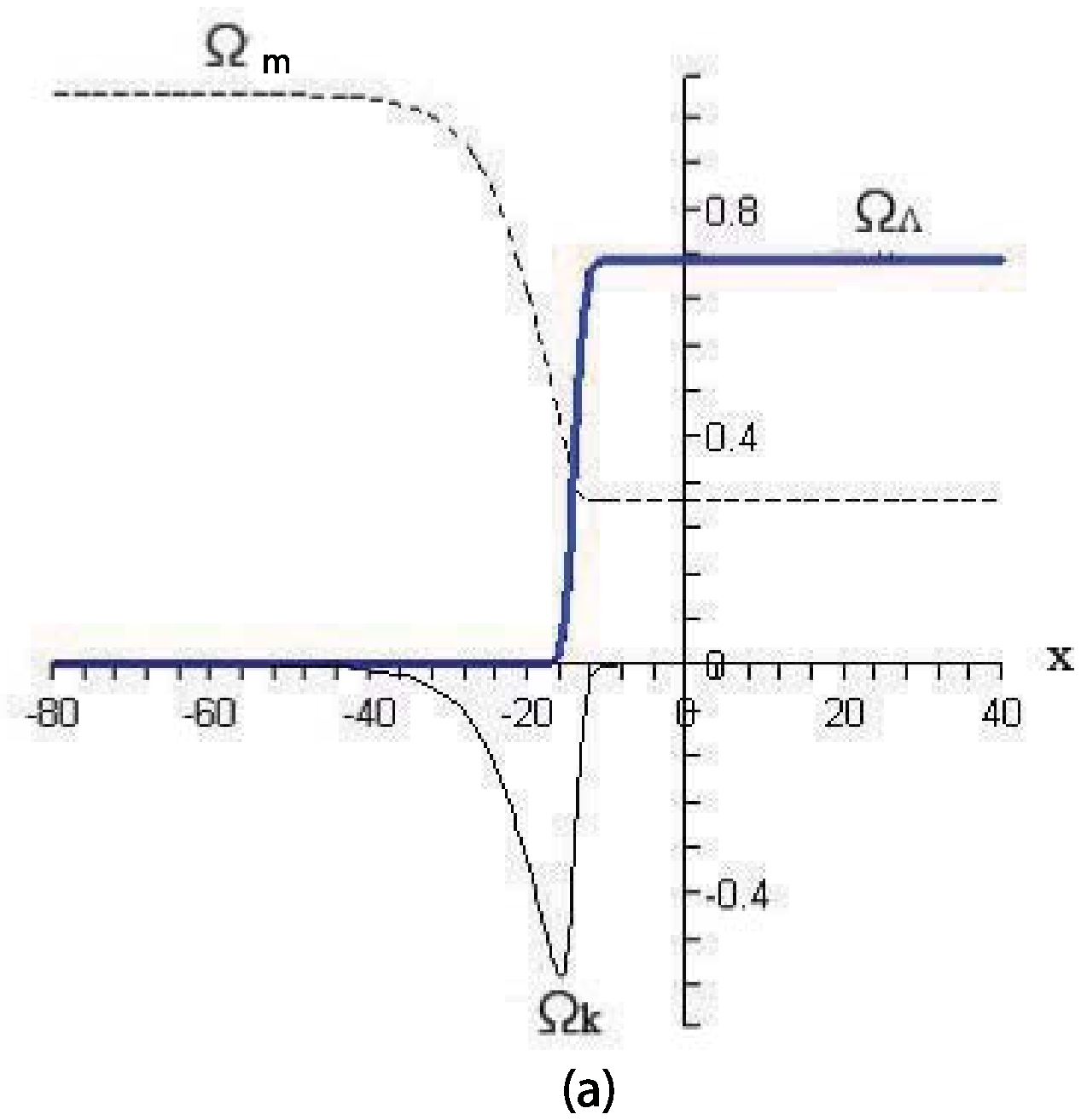}
\includegraphics[width=2.50in]{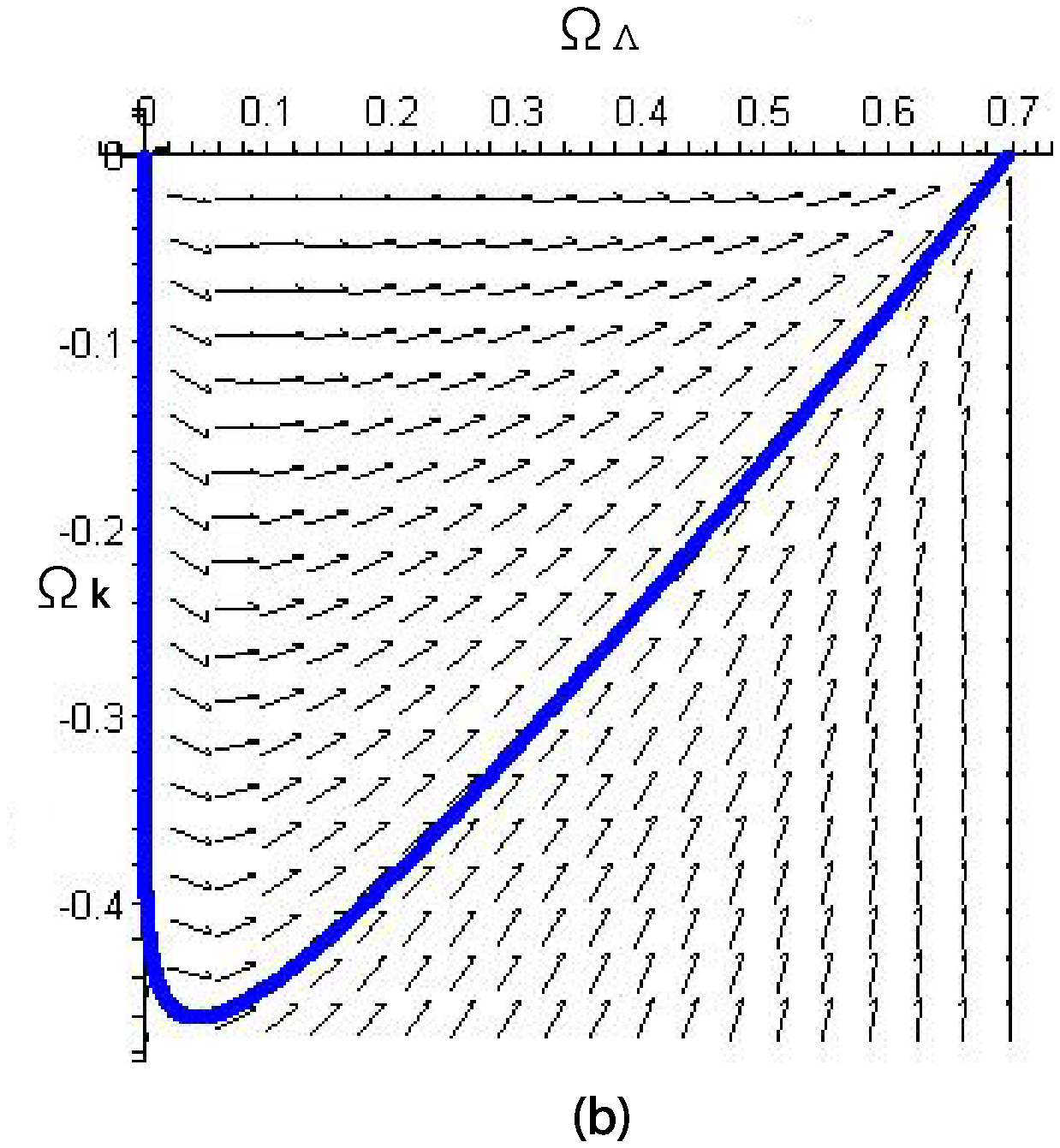}
\includegraphics[width=2.50in]{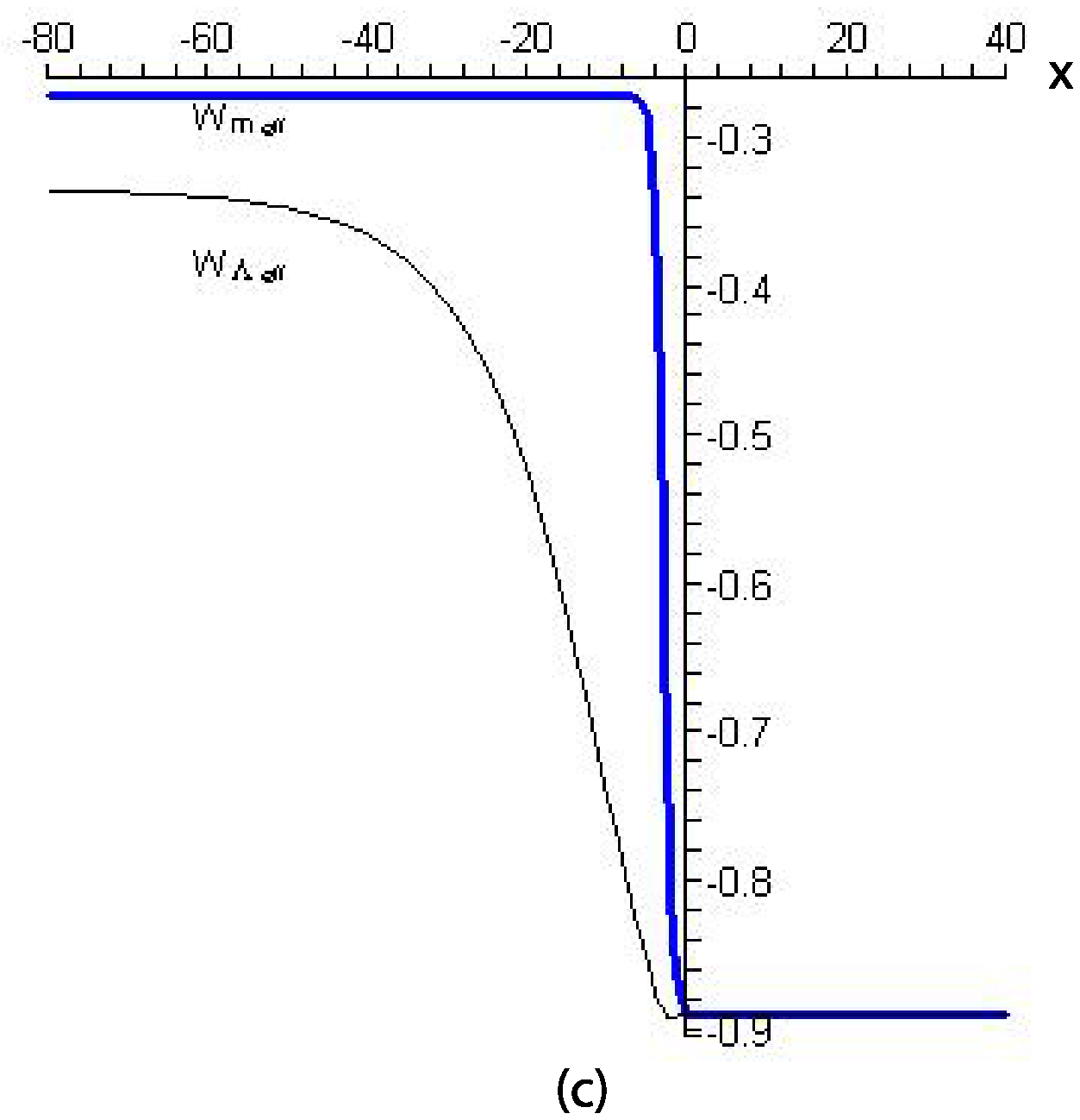}
}
}
\caption{(a) Density parameter evolution for the interaction
in Eq.~(\ref{usefulinteraction}), when $b^2=0.26$. (b) Flow diagram for the evoloution of density parameters using interaction Eq.~(\ref{usefulinteraction}). The solid line represents the solution depicted in (a). (c) Effective equations of state for dark energy and dark matter.
}
\label{graph5}
\end{figure}
\bigskip
\bigskip

In Fig.~\ref{graph5}, the Universe, is taken to be
filled with matter and to have very small curvature initially. It
moves toward a saddle point where the curvature is almost $-1$ and then evolves toward its final situation, where $\Omega_\Lambda
\approx 0.7$. How close the Universe approaches the intermediate saddle point depends on the initial values for
$\Omega_\Lambda$ and $\Omega_k$. If those initial conditions are chosen to
be smaller than the ones chosen in Fig.~\ref{graph5}, the evolution turns around
at smaller absolute values of curvature density parameter and then proceeds
toward its asymptotic solution. For the last fixed point, the effective equations of states are equal, and the
matter and dark energy density parameters remain at their
equilibrium solutions. Any curvature is eliminated as the
Universe expands under the influence of the presence of the dark energy component. This generalizes our previous result where an exactly flat initial
condition was assumed~\cite{Berger:2006db}.

\begin{figure}
\centerline{
 \mbox{\includegraphics[width=3.50in]{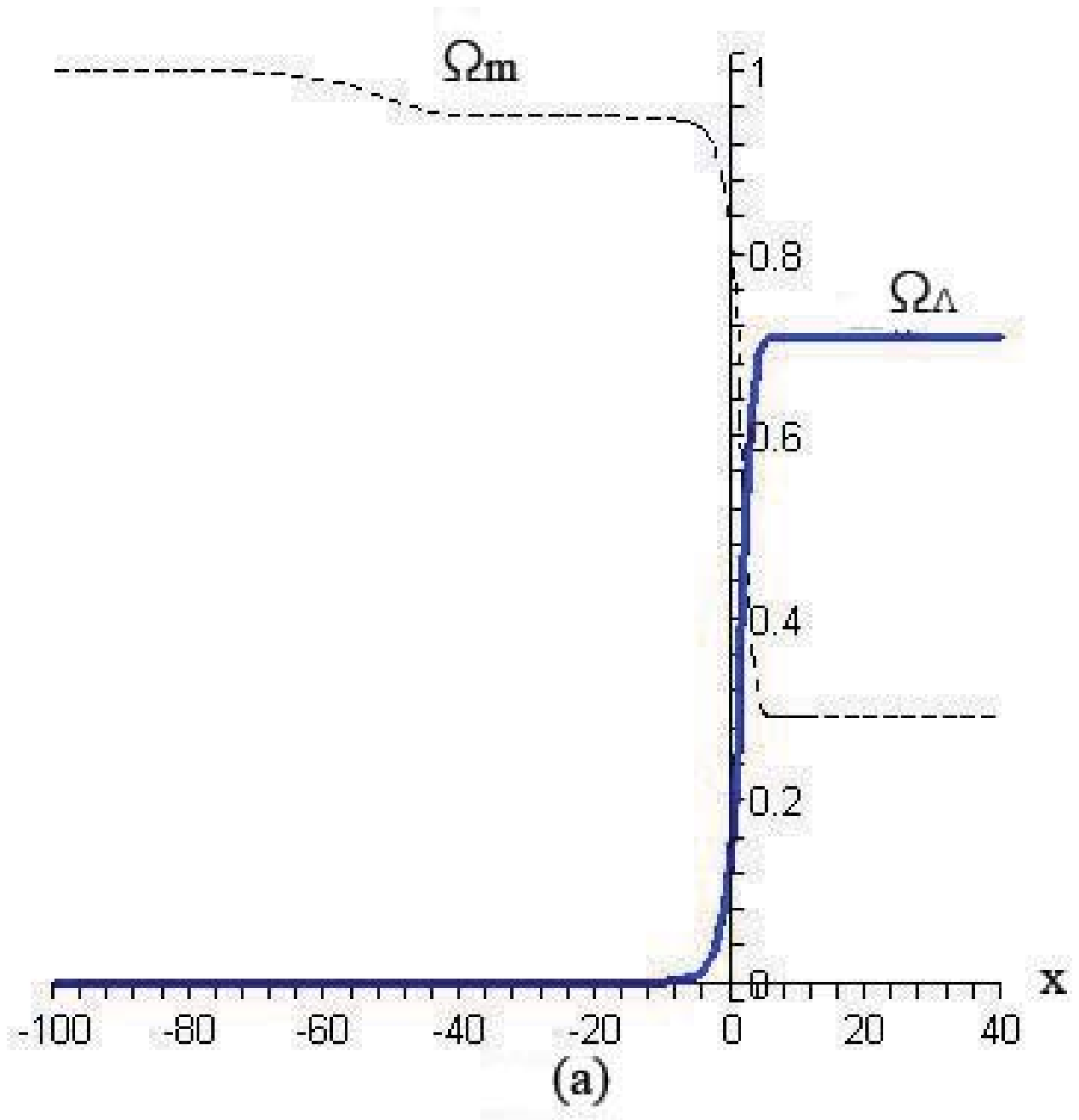}
\includegraphics[width=3.50in]{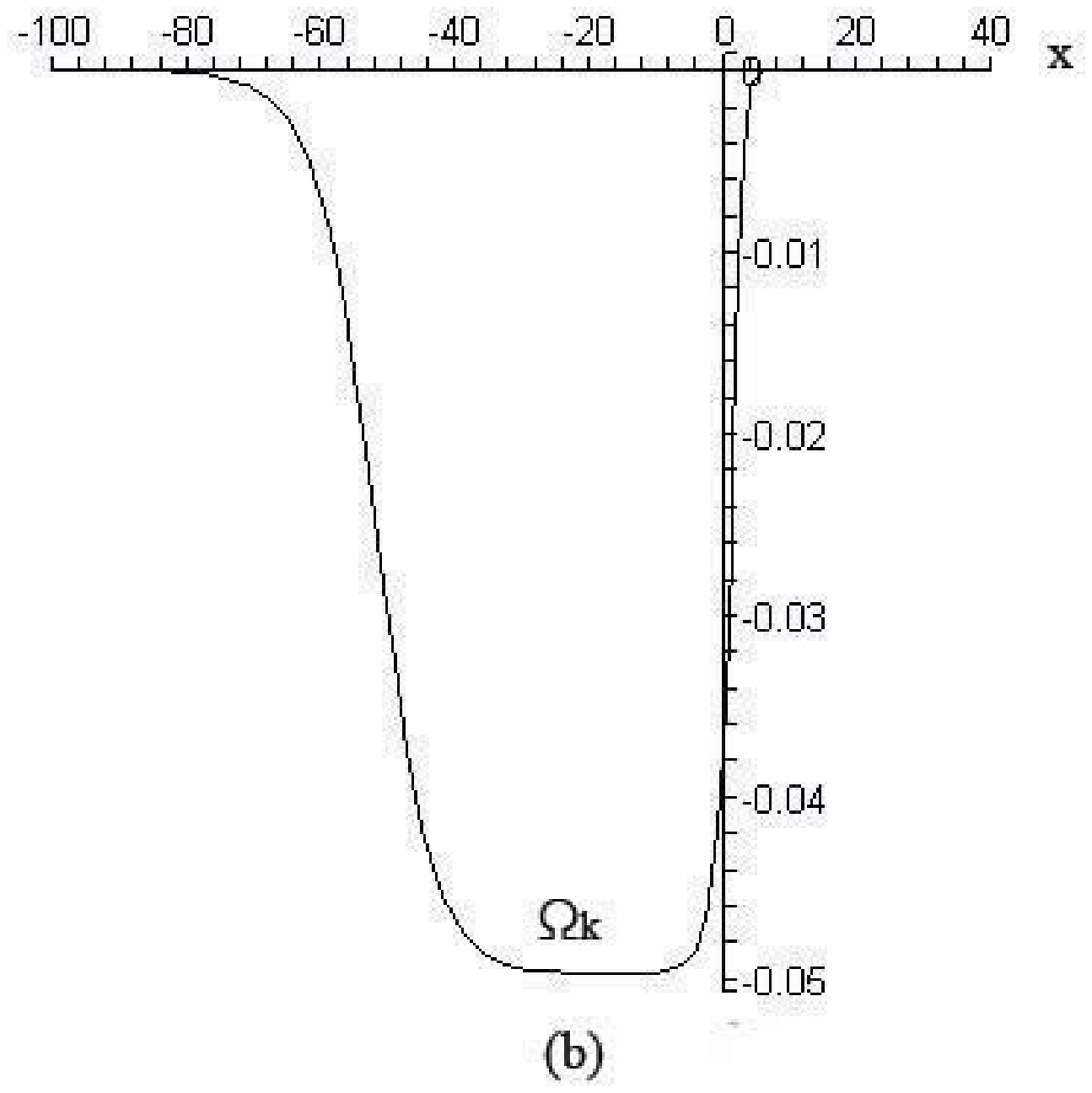}}
}
\caption{(a) Density parameters in an interacting model when $\Gamma$ is
proportional to $e^{-p\Omega_k}$,
when $b^{2}=0.26$ and $p=5$. (b) The initial value of the curvature
density parameter is taken to be a very small and negative number.}
\label{graph7}
\end{figure}

\begin{figure}
\centerline{
 \mbox{\includegraphics[width=2.50in]{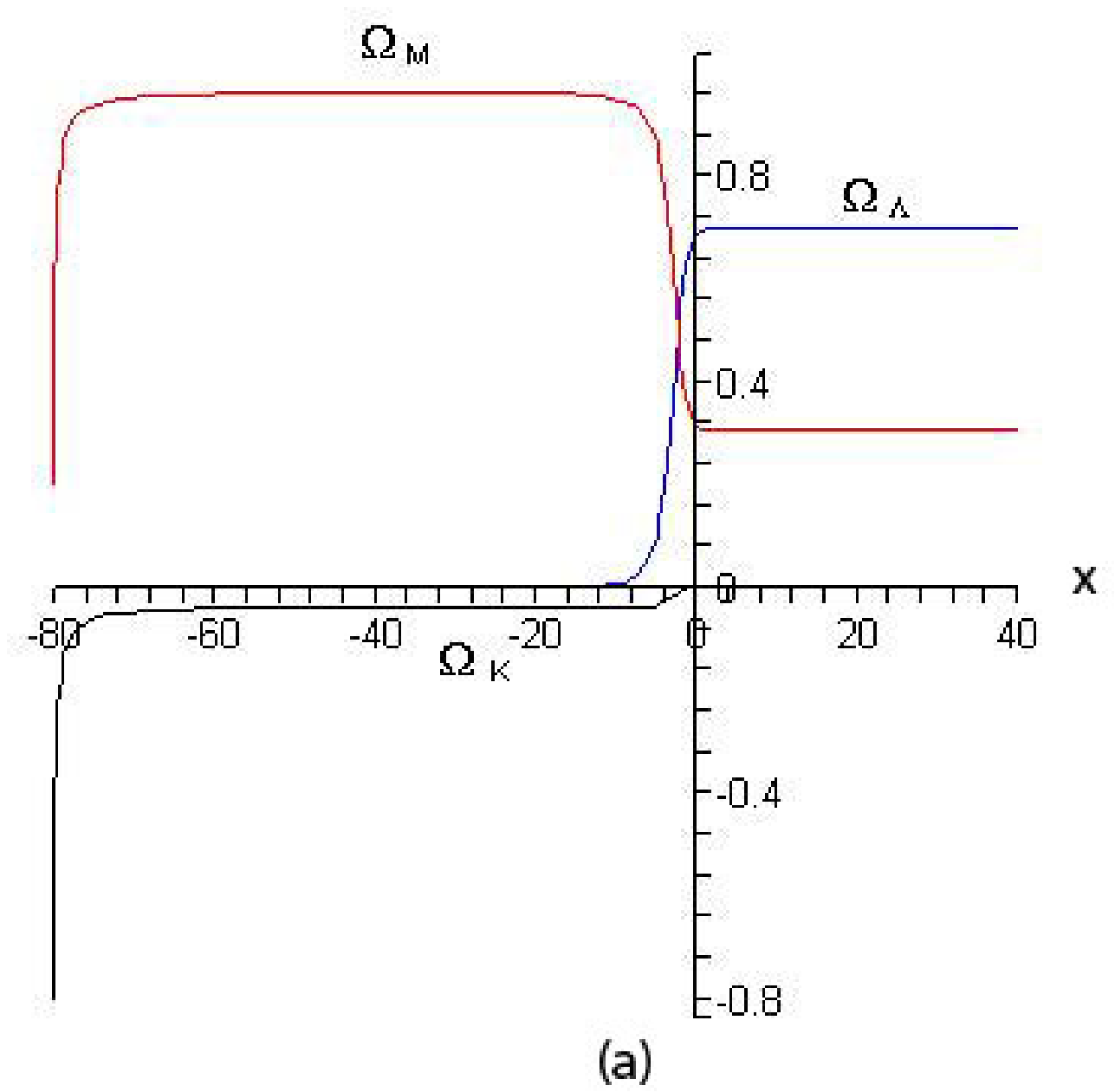}
\includegraphics[width=2.50in]{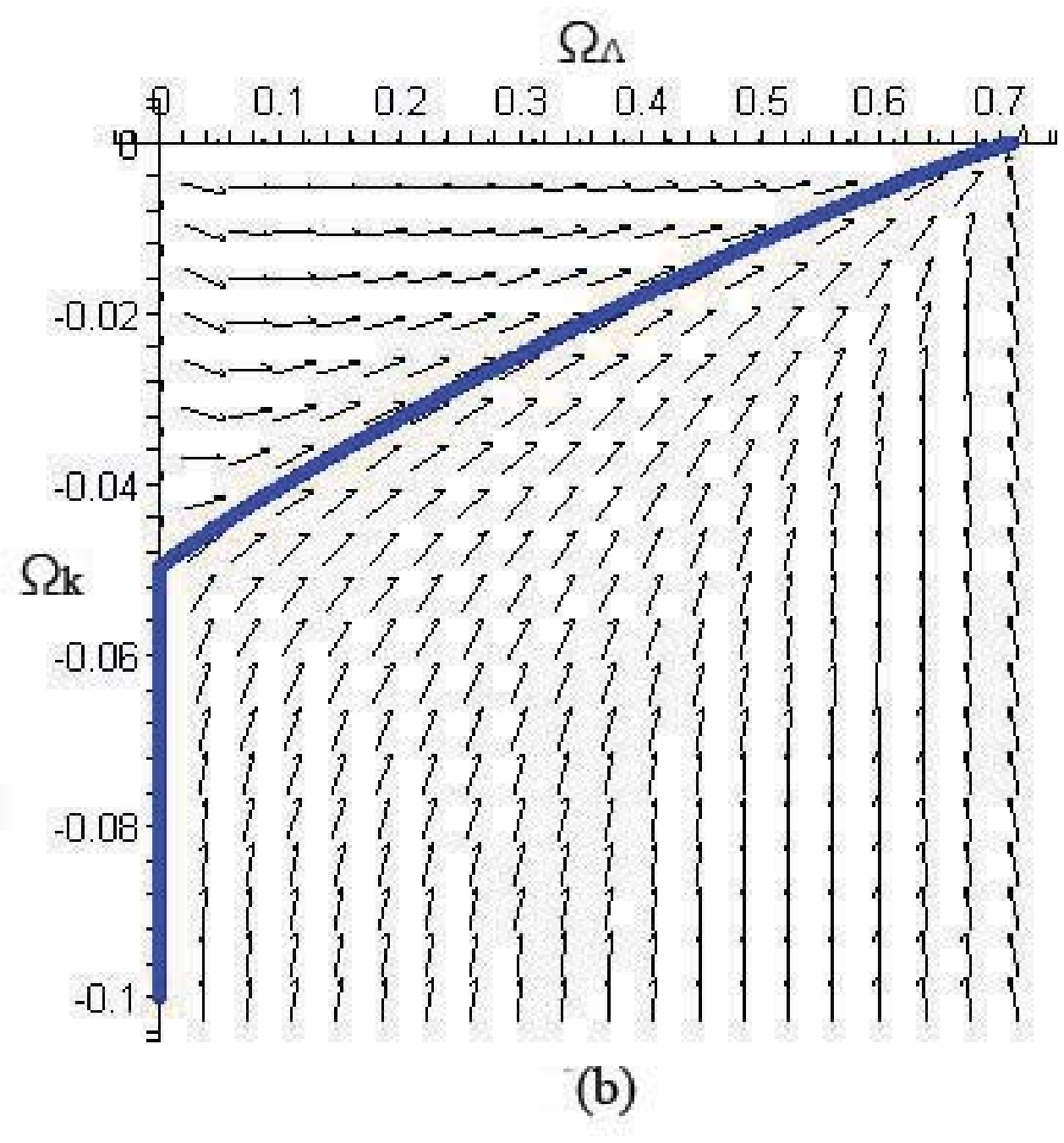}
\includegraphics[width=2.50in]{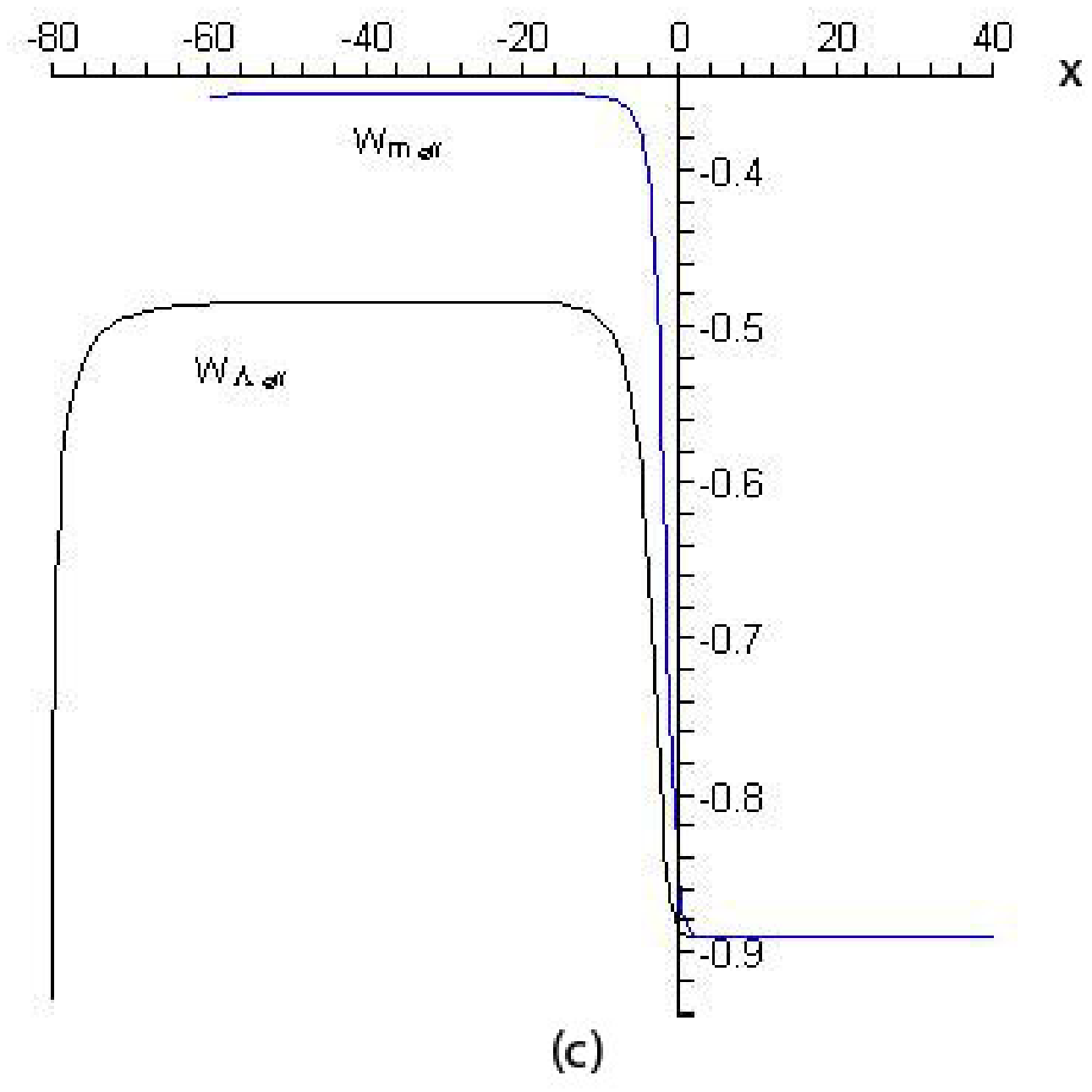}}
}
\caption{(a) Density parameters in an interacting model with
interaction, Eq.~(\ref{generalinteraction}). The
initial value of $\Omega_k$ is set close to $-1$. The
interaction drives $\Omega_k$ toward zero very rapidly. (b) The associated
flow diagram in the $\Omega_\Lambda,\Omega _k$ plane.
The solid line represents the solution depicted in (a),
for the region $\Omega_k>-0.1$. (c) Effective equations of state for dark energy and dark matter.}
\label{graph9}
\end{figure}

A final example, involves an interaction which causes the
universe to explore a transient state for a long period of time by
exploiting a fixed point which is a saddle point. Weighting the interaction, Eq.~(\ref{usefulinteraction}), by a factor of $\exp(-p\Omega_k)$,
we have
 \bea
{{\Gamma}\over{3H}}={{b^2e^{-p\Omega_k}(1+\Omega_k)}\over{\Omega_\Lambda}^n}\;.
\label{generalinteraction}
\eea
The exponential factor, while somewhat artificial, guarantees that evolution
is driven close to a saddle point.
The effective equation of state for matter is
\bea
w_m^{\rm
  eff}={{-b^2e^{-p\Omega_k}(1+\Omega_k)\Omega_\Lambda^{1-n}}
  \over{1+\Omega_k-\Omega_\Lambda}}\;.
\eea
We do not ascribe any physical motivation for such an interaction, but merely
employ it to show how the evolution can be driven close to a saddle point.
We again choose $n=1$ for the numerical solutions. In this case, the
Universe starts from a repeller, approaches the saddle point,
then is finally repelled toward
the attractor.
In Fig.~\ref{graph7} the initial value of the
curvature is set very close to zero and negative, while in
Fig.~\ref{graph9}, the initial value of the curvature density
parameter is set very close to $-1$. In
both cases, however, the Universe is pushed by the interaction toward
the saddle point where we have $\Omega_k\approx-0.05$ \footnote{This
  value depends on the specified values for $b^{2}$ and $p$ in the
  interaction.}, then goes toward the attractor
located at ${\bf \Omega}=(0.7,0)$. At this point, again the condition $w_m^{\rm eff}=w_\Lambda^{\rm eff}$ holds, as can be seen in Fig.~\ref{graph9} (c).

\bigskip
\bigskip

\bigskip
\bigskip
\bigskip

\section{conclusion}

The behavior of the Universe, where there is an interaction between dark
energy and dark matter, has been studied in the presence of curvature. In the literature, many different scenarios, obtained by choosing different interactions and different choices for length scale which enters the definition of dark energy density, have been explored. The present work unifies these attempts in this approach toward solving cosmic coinsidence problem, at the qualitative level, by studying the possible equilibria showing up in the dynamical equations of dark energy and dark matter.

Understanding the fixed point equilibria is important, because the asymptotic behavior at these points is insensitive to initial conditions. Attempts to solve the coincidence problem in the literature usually focus on stable solutions for which $\Omega_k=0$ and $w_m^{\rm eff}=w_\Lambda^{\rm eff}$. With appropriate choice of
interaction, a fixed point can occur at $\Omega_\Lambda\approx0.7$ and $\Omega_m\approx0.3$, the observed values for these parameters. We showed that adding curvature to cosmic inventory results in additional equilibium solutions. For example, the equilibrium point $\bf\Omega$ = (0,$-1$), which corresponds to $\Omega_m=0$, doesn't exist when there is no curvature. The new and old equilibrium solutions, and the possible behavior of them, are studied and categorized in Section III. A particular theory with an interaction between dark energy and dark matter doesn't typically possess all of these equilibria. For example if there is no interaction at all, there exists just the three equilibria listed in Table I. By a suitable choice of interaction, one is able to eliminate some of the equilibria and add new ones. The examples provided in Section IV illustrate how this process works.

Observation indicates that there is little curvature at the present epoch. Additional equilibria can be obtained by considering the Universe at all eras, where curvature is not necessarily small. Any saddle point equilibrium could represent a transient era (but perhaps long-lived) of Universe. Any attractor equilibrium, providing initial conditions let Universe be driven toward it, would represent the final state. In fact, no observation has confirmed that Universe is in a stable solution of its dynamical equations. So it is quite possible that Universe is currently experiencing a saddle point-like equilibrium and will ultimately be deriven toward a fixed point with complete different properties.

 The supernovae data requires a recent transition
from deceleration to acceleration, so a consistent solution requires that
we are arriving at this equilibrium at the present epoch. The fixed point solutions represent an amelioration of the rapid transition to dark energy
domination such as occurs in the $\Lambda$CDM. Any particular choice of interaction could be supported or ruled out by comparing its results with recorded history of Universe.  The examples provide insight into the classification of the equilibrium points.

\section*{Acknowledgments}
%\begin{theacknowledgments}
This work was supported in part by the U.S.
Department of Energy under Grant No.~DE-FG02-91ER40661.
%\end{theacknowledgments}

\end{document}